# GSA-DenseNet121-COVID-19: a Hybrid Deep Learning Architecture for the Diagnosis of COVID-19 Disease based on Gravitational Search Optimization Algorithm

**Dalia Ezzat[1,*], Aboul ell Hassanien[1,*], Hassan Aboul Ella[2,*]**

[1]Faculty of Computers and Artificial Intelligence, Cairo University, Egypt
[3]Microbiology Department, Faculty of Veterinary Medicine, Cairo University, Egypt
[*]Scientific Research Group in Egypt (SRGE) www.egyptscience.net
aboitcairo@cu.edu.eg

**Abstract**. In this paper, a novel approach called GSA-DenseNet121-COVID-19 based on a hybrid convolutional neural network (CNN) architecture is proposed using an optimization algorithm. The CNN architecture that was used is called DenseNet121 and the optimization algorithm that was used is called the gravitational search algorithm (GSA). The GSA is adapted to determine the best values for the hyperparameters of the DenseNet121 architecture, and to achieve a high level of accuracy in diagnosing COVID-19 disease through chest x-ray image analysis. The obtained results showed that the proposed approach was able to correctly classify 98% of the test set. To test the efficacy of the GSA in setting the optimum values for the hyperparameters of DenseNet121, it was compared to another optimization algorithm called social ski driver (SSD). The comparison results demonstrated the efficacy of the proposed GSA-DenseNet121-COVID-19 and its ability to better diagnose COVID-19 disease than the SSD-DenseNet121 as the second was able to diagnose only 94% of the test set. As well as, the proposed approach was compared to an approach based on a CNN architecture called Inception-v3 and the manual search method for determining the values of the hyperparameters. The results of the comparison showed that the GSA-DenseNet121 was able to beat the other approach, as the second was able to classify only 95% of the test set samples.

**Keywords**:. SARS-CoV-2, Deep Learning, Convolutional Neural Networks, Transfer Learning, Gravitational Search Algorithm, Hyperparameters Optimization

## 1. Introduction

On 11/March/2020, the world health organization (WHO) announced that the novel coronavirus disease-2019 (COVID-19) has been a Pandemic outbreak- COVID-19 is a respiratory disease caused by severe acute respiratory syndrome coronavirus 2 (SARS-CoV-2) which is a virus belongs to Coronavirdeae family to which both severe acute respiratory syndrome coronavirus 1 (SARS-CoV-1), the causative agent of 2002 severe acute respiratory syndrome (SARS) epidemic and middle east respiratory syndrome coronavirus (MERS-CoV), the causative agent of 2012 middle east respiratory syndrome (MERS) epidemic belong- This announcement was the beginning of the current medical health problem faced and shared by the whole world during the few last months. Up till now there are no efficient protective vaccines, neutralizing antisera or curative medication have been developed or officially approved to be used in COVID-19's patients worldwide.

The continuous increase in numbers of morbidities and mortalities due to severe acute respiratory syndrome coronavirus 2 (SARS-CoV-2) lead to international medical health worsen situation day after day. Therefore, this emerging COVID-19 pandemic becomes the ongoing challenge for all medical health workers and researchers.

By applying the natural timeline of infectious diseases on COVID-19 as shown in Figure (1), the importance of shortening the period between the onset of symptoms and usual time of diagnosis will appear. Therefore an efficient rapid diagnostic test or protocol will help to achieve proper early medical caring to COVID-19 patients that by its role will help to save a lot of lives worldwide. Finding a rapid efficient diagnostic test or protocol becomes one of those top critical priorities.

Quantitative reverse transcriptase polymerase chain reaction (qRT-PCR) is the golden standard test for confirmed laboratory diagnosis of COVID-19. Other rapid, bedside, field and point of care immunochromatographic lateral flow, nucleic acid lateral flow, nucleic immunochromatographic lateral flow and CRISPR *cas*-12 lateral flow are under developing.

COVID-19 as a pneumonic disease characterized by general pneumonic lung affections with certain uniqueness from other pneumonia causing Coronaviruses. Although the radiological imaging features closely similar and overlapping those associating of SARS and MERS, the bilateral lungs involvement on initial imaging is more likely to be seen with COVID-19; as those associating (SARS) and (MERS) are more predominantly unilateral. So using radiological imaging techniques as X-rays and computed tomography (CT) is of great value as confirmed, need expert but rapid diagnostic technique either separately or in combination with quantities reverse transcriptase polymerase chain reaction (qRT-PCR) to avoid the false positive COVID-19 results which have been recorded and reported during separate use of polymerase chain reaction (PCR) in early stage of the disease.

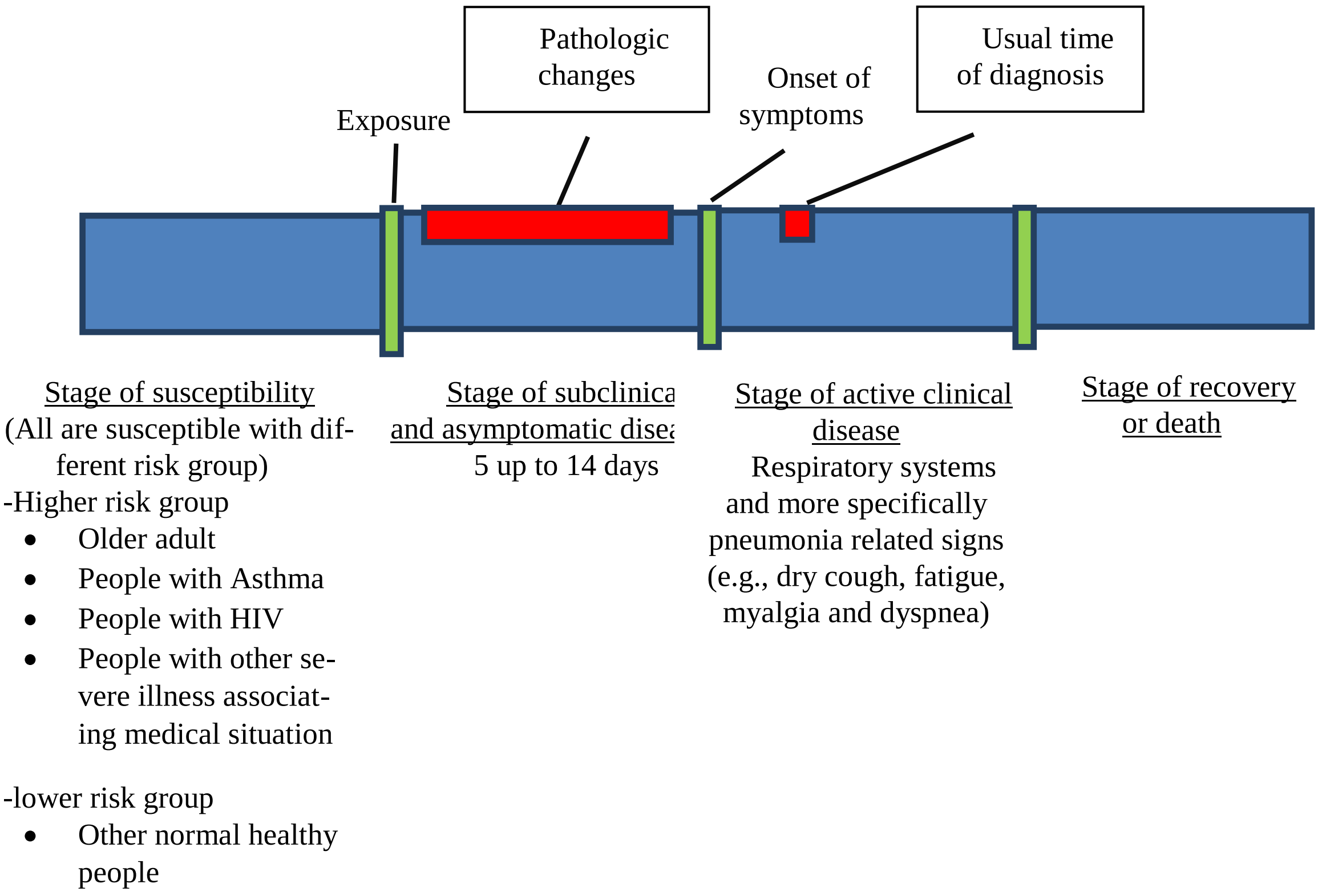

**Figure (1)** Natural timeline of infectious diseases on COVID-19

In this research article, CNN architecture is adapted, which is the preferred DL architecture for diagnosis of COVID-19 through chest radiological imaging, especially X-rays.

Deep learning (DL) is the most common and accurate method for dealing with medical data sets such as the classification of brain abnormalities, the classification of different types of cancer,  the classification of pathogenic bacteria, and biomedical image segmentation [1-5]. DL is a kind of machine learning (ML) methods based on artificial neural networks. It has significantly improved the performance of artificial intelligence tasks such as image classification, machine translation, and many other tasks. The deep structure nature of the DL architectures gives the DL the ability to solve many of the most complex artificial intelligence tasks. As a result, DL has gained wide attention in both academia and industry [6]. The DL is exciting and this is due to several reasons, the most important of which is that its performance level is greater than other ML methods. As well as, it does not require human intervention to extract and identify important features. During the training phase, the DL architectures learn the features that contribute to achieving the best results themselves, which means that no engineering advantage is required [7]. However, unlike ML methods, DL architectures such as convolutional neural networks (CNNs) and recurrent neural networks (RNNs) contain many hyperparameters. The values of these hyperparameters are not learned during training the network, and therefore, the user must choose their values before starting the training phase [8].

Many studies such as [9, 10] conducted to find out the extent of the influence of hyperparameters on various DL architectures. These studies have found the hyperparameters that offer significant performance improvements in simple networks do not have the same effect in more complex networks. As well as, the hyperparameters that fit one dataset do not fit another dataset with different properties. With no mathematical formula to choose the appropriate values for hyperparameters, their choice often depends on a combination of human experience, trial and error or the use of a grid search method [11].  Due to the nature of computationally expensive DL algorithms, which can take several days to train, the trial and error method is ineffective [12]. The grid search method is usually not suitable for DL architectures, because the number of combinations grows exponentially with the number of hyperparameters. Therefore, automatic optimization of hyperparameters of deep learning algorithms is so important. Recently, the selecting of the hyperparameters values has been formulated as an optimization problem by the researchers such as [13,14].

In this paper, a CNN architecture is used, which is the preferred DL architecture for most image analysis tasks such as image classification. In order to obtain the best performance of the CNN, the method of transfer learning was applied. In order to choose the optimal values for the hyperparameters of the used CNN architecture and improve its performance, the gravitation search algorithm (GSA) [15] is used.

The main contribution of this paper can be summarized in the following points:

- The dataset used was collected from two different datasets to build an approach that can diagnose cases with COVID-19 and differentiate them from other cases that are similar in clinical symptoms but are infected with other types of viruses or bacteria.
- The proposed approach for diagnosing the COVID-19 virus involves a hybrid CNN architecture using the GSA.
- The results of this proposed approach to the diagnosis of COVID-19 virus were very effective, achieving a 98% accuracy level on the test set.

The rest of the paper is structured as follows: Section 2 represents the theoretical background for the CNNs and the GSA. The dataset used is discussed in Section 3. Section 4 shows details of the proposed approach. The results achieved by the proposed approach are illustrated in Section 5.

## 2. Theory and Methods

### *2.1 Gravitational Search Algorithm*

GSA is an optimization technique that has been gaining attention in the last years and developed by Rashedi [15]. It is based on the law of gravity, as shown in Equation (1) and the second law of motion as shown in Equation (3) [16]. It also depends on the general physical concept, that there are three types of mass: inertial mass, active gravitational mass and passive gravitational mass [17]. The law of gravity states that every particle attracts every other particle with gravitational forces (F). The gravitational forces (F) between two particles is directly proportional to the product of their masses ($M_1$ and $M_2$) and inversely proportional to square of their distance ($R^2$). The second law of motion states that when a force ($F$) is utilized to a particle, its acceleration ($a$) is determined by the force and its mass ($M$).

$$\mathrm{F} = \mathrm{G}\frac{\mathrm{M_1 M_2}}{\mathrm{R^2}} \quad (1)$$

Where $G$ is the gravitational constant which decreases with increasing time and it is calculated as equation (2) [18].

$$G(t) = G(t_0) \times \left(\frac{t_0}{t}\right)^{\beta} \quad , \beta < 1 \quad (2)$$

$$\mathrm{a} = \frac{\mathrm{F}}{\mathrm{M}} \quad (3)$$

The GSA is similar to the above basic laws with minor modifications in equation 1, where Rashid [15] stated that based on the experimental results, inverse proportionality to distance (R) produces better results than $\mathrm{R}^2$.The GSA can be expressed as an isolated system of $N$ particles and their performance is measured by their masses. All of these particles attract each other by the force of gravity, and this force causes a universal movement of all particles towards the particles that have heavier mass. Consequently, masses collaborate using a direct form of communication, through the force of gravity. The heavy masses represent good solutions as they move more slowly than lighter mass, while light masses represent worse solutions, moving towards the heavier masses faster. Each mass has four specifications: active gravitational mass, passive gravitational mass, inertial mass, and position. The mass' position corresponds to a solution of the problem and the other specifications of the mass (active gravitational mass, passive gravitational, inertial mass) are determined utilizing the fitness function. The algorithm of GSA can be summarized in eight steps as follows:

- ***Step one: Initialization***

Assuming there is an isolated system with N particles (masses), the position of $i$th particle is denoted as :

$$P_i = \left(p_i^1, p_i^2, \ldots, p_i^d, \ldots, p_i^N\right) \quad \text{for } i = 1,2,3, \ldots, \mathrm{N} \quad (4)$$

Where $p_i^d$ presents the position of $i$th particle in the $d$th dimension.

- ***Step two: Fitness evaluation of particles***

In this step the worst and best fitness are calculated as equations (5) and (6) respectively for a minimization problem, and calculated as equations (7) and (8) respectively for a maximization problem.

$$worst(t) = \max_{j\epsilon\{1,\ldots,N\}} \text{fitness}_j(t) \tag{5}$$

$$best(t) = \min_{j\epsilon\{1,\ldots,N\}} \text{fitness}_j(t) \tag{6}$$

$$worst(t) = \min_{j\epsilon\{1,\ldots,N\}} \text{fitness}_j(t) \tag{7}$$

$$best(t) = \max_{j\epsilon\{1,\ldots,N\}} \text{fitness}_j(t) \tag{8}$$

Where $\text{fitness}_j(t)$ is the fitness of the $j$th particle at time $t$.

- ***Step three: Calculate the gravitational constant $G(t)$***

In this step, the gravitational constant $G(t)$ at time $t$ is calculated as follows [19]:

$$G(t) = G_0 \times (1 - \frac{t}{t_{\max}}) \tag{9}$$

Where $G_0$ represents the initial value of the gravitational constant initialize randomly, $t$ is the current time, $t_{\max}$ is the total time.

- ***Step four: Update the inertial and gravitational masses***

In this step, the inertia and gravitational masses are updated by the fitness function. Assuming the equality of the inertia and gravitational mass, the masses' values are calculated as follows:

$$M_{ii} = M_{pi} = M_{ai} = M_i \qquad \text{for } i = 1,2,3,\ldots,N \tag{10}$$

$$m_i(t) = \frac{\text{fitness}_i(t) - worst(t)}{best(t) - worst(t)} \tag{11}$$

$$M_i(t) = \frac{m_i(t)}{\sum_{j=1}^{N} m_j(t)} \quad (12)$$

Where $\text{fitness}_i(t)$ is the fitness of the $i$th particle at time $t$, $M_i(t)$ is the mass of the $i$th particle at time $t$.

- ***Step five: Compute the total force***

  In this step the total force $F_i^d(t)$ that exerting on particle $i$ in a dimension $d$ at time $t$ is calculated as follows:

$$F_i^d(t) = \sum_{j=1, j\neq i}^{kbest} rand_j F_{ij}^d(t) \quad (13)$$

Where $rand_j$ is a random number $\in [0,1]$, $kbest$ is the set of first $kbest$ particles with the best fitness value and the biggest masses. $F_{ij}^d(t)$ is the force exerting from mass '$j$' on mass '$i$' at time '$t$' and it calculated as the following equation:

$$F_{ij}^d(t) = G(t) \frac{M_{pi}(t) \times M_{aj}(t)}{R_{ij}(t) + \tau} \left(p_j^d(t) - p_i^d(t)\right) \quad (14)$$

Where $M_{pi}$ represents the passive gravitational mass associated with particle $i$, $M_{aj}$ represents the active gravitational mass associated with particle $j$. $\tau$ is a small positive constant to prevent division by zero, $R_{ij}(t)$ represents the Euclidian distance between particles $j$ and $i$:

$$R_{ij}(t) = \left\| p_i(t), p_j(t) \right\|_2 \quad (15)$$

- ***Step six: compute the velocity and acceleration***

  In this step based on $F_i^d(t)$, the acceleration of the particle $i$ , $a_i^d(t)$, at time $t$ in the direction $d$th, and the next velocity of the particle $i$ in the direction $d$th , $u_i^d(t+1)$, are calculated as follows:

$$a_i^d(t) = \frac{F_i^d(t)}{M_{ii}(t)} \quad (16)$$

$$u_i^d(t+1) = rand_i \times u_i^d(t) + a_i^d(t) \tag{17}$$

Where $M_{ii}(t)$ represents the inertial mass of $i$th particle, $rand_i$ is a random number $\in [0,1]$.

- ***Step seven: update particles' position***

  In this step the next position of the particle $i$ in the direction $d$th, $p_i^d(t+1)$, is calculated as follows:

$$p_i^d(t+1) = p_i^d(t) + u_i^d(t+1) \tag{18}$$

- ***Step eight:*** repeat steps two to seven until the stop criteria are reached, these eight steps are simpler illustrated by a flowchart shown in Figure 2.

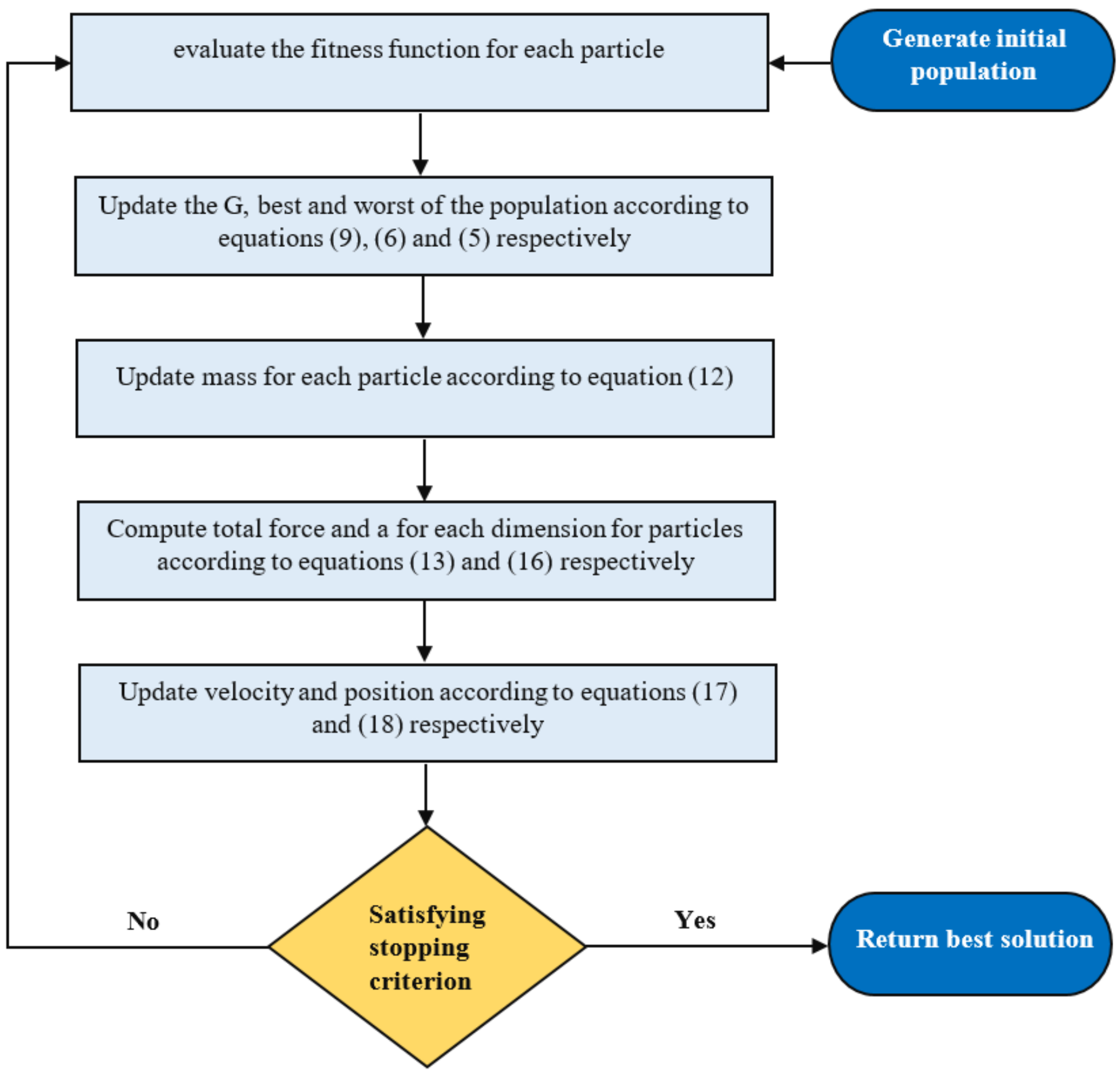

**Figure (2)** GSA flowchart

### *2.2 Convolutional Neural Networks*

**Network Structure:** CNN architectures consist of two bases namely convolutional base and classifier base. The convolutional base includes three major types of layers are: a convolutional layer, an activation layer, and a pooling layer, utilized to discover the critical features of the input images, called feature maps. While the classifier base includes the dense layers that convert the feature maps to one dimension vectors to expedite the classification task using a number of neurons. In this section, some of the most frequently used layers to build CNN architecture will be described. As well as, dropout, which is an important trick, will be explained in any DL architecture.

**(1) Convolutional Layers:** These layers produce several feature maps. A feature map is getting by performing convolution processes to the input image or prior features using a linear filter, merging a bias term, and then passing this feature map through a non-linear activation function. In other words, each neuron in a feature map receives inputs from the combination of an N ×N region of a subset of all of the features maps of prior layers or a subset of the input layer. The combined regions are known as the receptive fields of this neuron. Each neuron in the same feature map share the same weights with the corresponding receptive field as the same filter in the convolutional layer is utilized to look into all bearable receptive fields of previous feature maps. The shared weights, also known as filters or kernels, are learned by during the training phase. Using multiple convolutional layers in the network can help the network to learn more important features, as the first convolutional layers detect simple features such as lines, and the last convolutional layers detect more abstract features such as nose and eye [20].

**(2) Activation Function Layers**: Activation functions are mathematical equations that decided the output of a neural network. The function is connected to every neuron within the network and decides whether or not it ought to be activated or not, based on whether or not every neuron's input has relevance for the prediction of the model. For each neuron, the input is multiplied by the weights in the neuron and then merged together. The result of this process is called the summary activation of the neuron and is then transformed using the activation function which in turn decides the outputs of the neuron [21]. There are two forms of the activation func-tions are linear activation functions and nonlinear activation functions. The linear activation functions are very simple and do not lead to any transformation to the outputs of the neurons. Nonlinear activation functions are preferred over linear activation functions because they permit neurons to detect more complicated information in any dataset. There are many types of nonlinear activation functions such as Sigmoid, Rectified Linear Unit.

The sigmoid activation function is a very common activation function for DL architectures. This function converts any value pass through it to a value between 0 and 1, where values greater than 1 are converted to 1 and values smaller than 0 are converted to 0. It is mathematically defined as equation (19) [22].

$$Sigmoid(z) = \frac{1}{1 + e^{-z}} \tag{19}$$

Rectified Linear Unit (RELU) is a very effective and simple activation function; it acts as a nonlinear activation function and linear activation function. As it returns the value provided as input without any transformation or returns 0 if the input value is 0 or less. It computationally defined as equation (20) [23].

$$RELU(z) = max(0, z) \tag{20}$$

**(3) Pooling Layers:** There is a restriction of the feature maps produced from the convolutional layers is that they register the exact placement of the features in the input image. This implies that a slight change in the feature's placement in the input image will lead to a completely different feature map. Changes can occur by applying data augmentation techniques such as cropping and rotating. A popular method to handle this problem is known as down-sampling, through this method a lower resolution version of the input image is generated that includes the important features only without the features that may not be beneficial to the task. A more powerful method is to utilize a pooling layer. The pooling layer is a layer that follows the convolutional layer that is activated by the nonlinear activation function. It is responsible for minimizing the number of values within the feature maps produced from the prior convolutional layer by specifying only the most important values in the feature maps. There are two common types of pooling are the maximum pooling and the average pooling. The maximum pooling takes the maximum worth for every patch within the feature map. The average pooling computes the average of values of every patch within the feature map [24].

**(4) Dropout:** It is an effective regularization technique designed to minimize the overfitting that may encounter DL architectures and improve their generalization. The dropout indicates that some neurons or units are dropped or temporarily removed from the network during the training stage, along with their connections to other units. This technique has the impact of making the training operation noisy, compelling the nodes within a layer to take more or less responsibility for inputs. Dropout is performed for the layers of the neural network architecture. It can be utilized with most forms of layers, such as dense layers, convolutional layers, however, it cannot be applied to the output layer. Dropout technique introduced a new hyperparameters called dropout rate, which determines the probability at which outputs of the layer are removed, or reciprocally, the probability at which outputs of the layer are kept. Typically, the dropout rate is set in the range from 0.1 to 0.9 [25, 26].

**Transfer Learning:** A popular and extremely efficient approach on a tiny image dataset is to utilize a pre-trained network. A pre-trained network is a network that was trained on a huge dataset, usually in the task of categorizing images, and then its architecture and weights were preserved. If this initial dataset is big enough and general enough, the amount of the features learned by the pre-trained network can it to be effective as a general model of the visual world. Therefore, these features can be helpful for several totally different computer vision tasks, even if the new tasks may contain fully totally different categories from categories of the initial task [27, 28]. For example, networks that have been trained on the ImageNet database, such as DenseNet121 [29], can reset to something as remote as exploring medical image features.

Transfer learning from a pre-trained network can be applied in two ways namely feature extraction and fine-tuning. Feature extraction involves taking the convolutional base of a pre-trained network to extract the features of the new dataset and then training a new classifier on top of this output. Fine-tuning is a complementary to feature extraction method, where it involves unfreezing the last layers of the frozen convolutional base utilized for the feature extraction, and retraining these layers jointly with the new classifier previously learned in feature extraction method. The purpose of the fine-tuning method is to adjust the most abstract features of the pre-trained model, to make them more relevant to the new task. The steps for using these methods can be explained as follows [30]:

- A pre-trained network is taken and its classifier base is removed.
- The convolutional base of the pre-trained model is frozen.
- A new classifier is added and trained on top of the convolutional base of the pre-trained network.
- Unfreeze some layers of the convolutional base of the pre-trained network.
- Finally, both these unfrozen layers and the new classifier are jointly trained.

**Performance Metrics:** Several performance metrics can be used to evaluate the performance of CNNs such as accuracy, error rate, precision, recall, F1-score, and confusion matrix. Accuracy is one among the foremost remarkably used measures for measuring the performance of classification models, and it is outlined as a proportion between the properly classified samples to the overall number of samples as shown in equation (21). The error rate is the complement of the accuracy, it represents the samples that misclassified by the model and calculated as equation (22) [31].

$$Accuracy = \frac{TP + TN}{TP + TN + FP + FN} \tag{21}$$

$$Error\ rate = 1 - Accuracy = \frac{FP + FN}{TP + TN + FP + FN} \tag{22}$$

Where *P*= the number of the positive samples, and *N*= the numbers of the negative samples.

Precision as shown in equation (23), it is the number of true positives divided by the number of true positives and false positives. In other words, it is the number of positive predictions divided by the total number of positive category values predicted. Precision can be considered a measure of the rigor of a classifier. A low precision can also indicate a large number of false positives [32].

$$Precision = \frac{TP}{TP + FP} \tag{23}$$

Recall, which also termed as sensitivity is the number of true positives divided by the number of true positives and the number of false negatives as shown in equation (24). In other words, it is the number of positive predictions divided by the number of positive class values in the test set. Recall can be considered a measure of how complete a classifier is. A low Recall indicates many false negatives [32].

$$Recall = \frac{TP}{TP + FN} \tag{24}$$

F1 Score, which also termed as F Score, is a function of precision and recall and calculated as equation (25). It is used to seek a balance between precision and recall [32].

$$F1\ Score = 2 * \frac{Precision * Recall}{Precision + Recall} \tag{25}$$

Confusion Matrix is a synopsis of the prediction results regarding the classification problem. The confusion matrix gives insight into not only the mistakes committed by the classifier, but more importantly the types of mistakes that are made [33].

## 3. Binary COVID-19 Dataset Description

The binary COVID-19 dataset used in this paper is a combination of two datasets, the first dataset is the COVID19 Chest X-ray dataset made available by Dr. Joseph Paul Cohen of the University of Montreal [34]. This dataset consists of 150 chests X-ray and CT images as of the time of writing this paper, 126 images of this dataset represent cases infected with the COVID-19 virus. While the rest of the images represent cases infected with other viruses such as SARS and other diseases such as acute respiratory distress syndrome (ARDS). The second dataset is the Kaggle Chest X-ray dataset made available for a Data Science competition [35]. This dataset consists of 5811 X-ray images, 1538 images represent normal cases and 4273 images represent pneumonia cases.

The binary COVID-19 dataset was built to distinguish COVID-19 cases from those suffering from other diseases and the healthy cases using only X-ray images where CT images were removed from the COVID19 Chest X-ray dataset. Therefore, the used dataset consists of two categories: positive and negative, the positive category contains 99 X-ray images representing cases infected with the COVID-19 virus from the COVID19 Chest X-ray dataset. The negative category contains 207 X-ray images, some from the COVID19 Chest X-ray dataset and some from the Kaggle Chest X-ray dataset. Some images of each category of the binary COVID-19 dataset are shown in Figure 3.

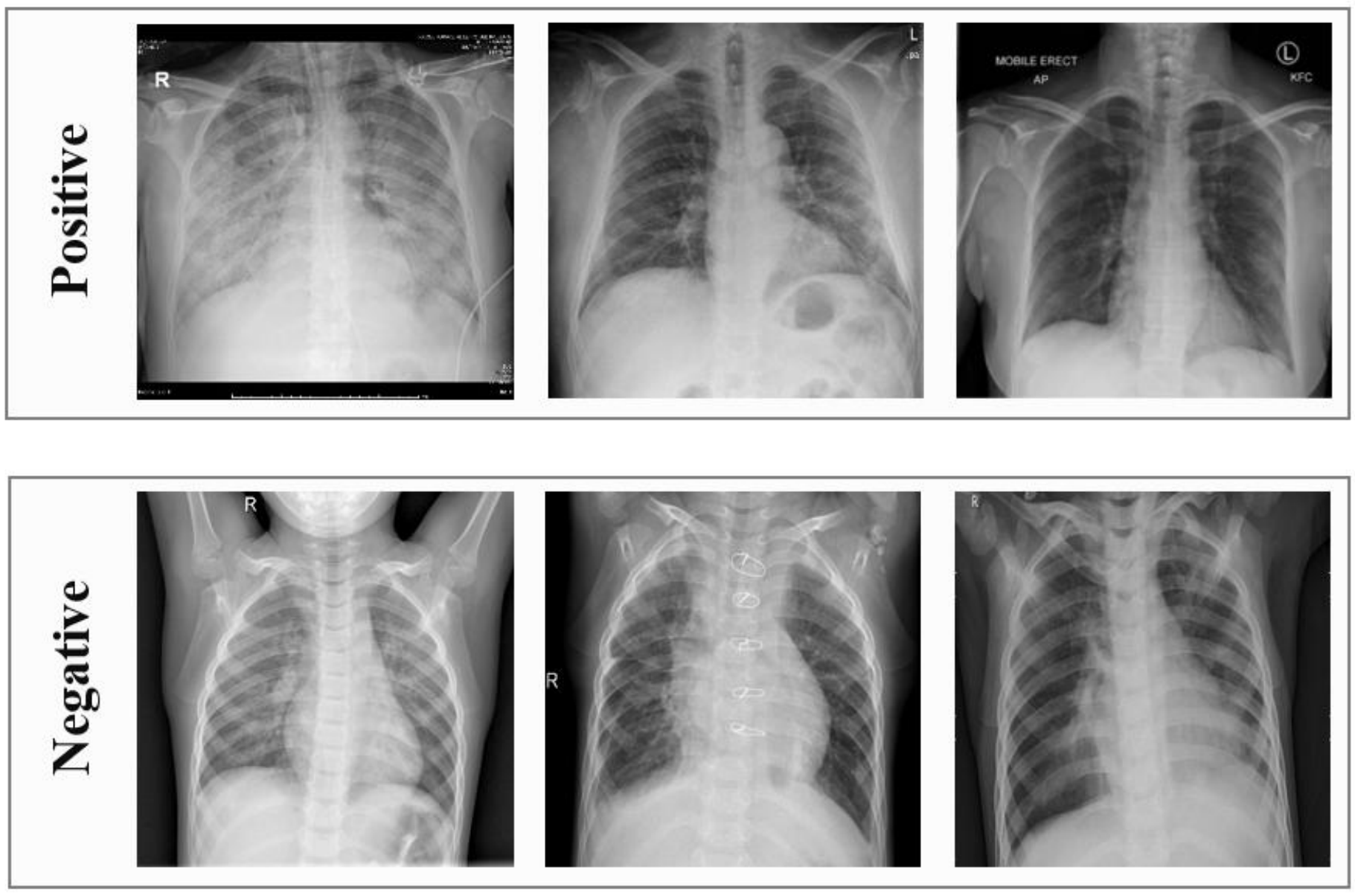

**Figure (3)** Some images of the positive and negative categories of the binary COVID-19 dataset

## 4. The Proposed GSA- DenseNet121-COVID-19 Approach

The proposed GSA- DenseNet121-COVID-19 approach relies on the transfer learning from a pre-trained CNN architecture. The pre-trained architecture utilized in the proposed approach is DenseNet121. For the best performance of this architecture, its hyperparameters that most influence its performance have been optimized using the GSA. After determining the optimal values for these hyperparameters, DenseNet121 was trained using transfer learning techniques. Once this training is completed, it is evaluated using a separate test set. In other words, the training and validation sets were used to determine the optimal values for the hyperparameters of the DenseNet121 and trained it, whereas the fully trained DenseNet121 is then evaluated using the test set. To facilitate the description of the proposed approach, it has been divided into four main stages as shown in Figure 4. The first stage is the data preparation, the second stage is the hyperparameters selection, the third stage is the learning, and the performance measurement is the fourth stage. Each stage will be explained in detail through the following section.

### *(A) Data Preparation Stage*

As explained in the data description section, the positive category of the binary COVID-19 dataset contains 99 samples, while the negative category contains 207 samples, which means that this dataset is not balanced. In most cases, not all ML algorithms can handle this type of dataset well. Because most of the information available in this type of dataset belongs to the dominant category, making any ML algorithm learn to categorize the dominant category and not categorize the other minor category. Therefore, the number of images in the positive category has been increased by randomly copying some images, however, after cutting each image so that random copying does not cause the used DL algorithm to overfit the dataset. After that, the data set became balanced, with each category containing 207 images.

The balanced binary COVID-19 dataset was divided into three sets: training set, validation set, and testing. The training set contains 70% of the dataset, that is, it contains about 146 images in each category, while each of the validation set and the test set contains 15% of the dataset samples, meaning each of them contains 31 images in each category. To reduce the overfitting and improve generalization, various data augmentation techniques [36] have been applied to increase the number of training samples. The data augmentation techniques used in this paper are: brightness, rotation, width shift, height shift, shearing, zooming, vertical flip, and horizontal flip. As well as feature wise centering, feature wise standard deviation normalization and fill mode. Before the images were supplied to other stages, they were resized to a resolution of 180 x 180.

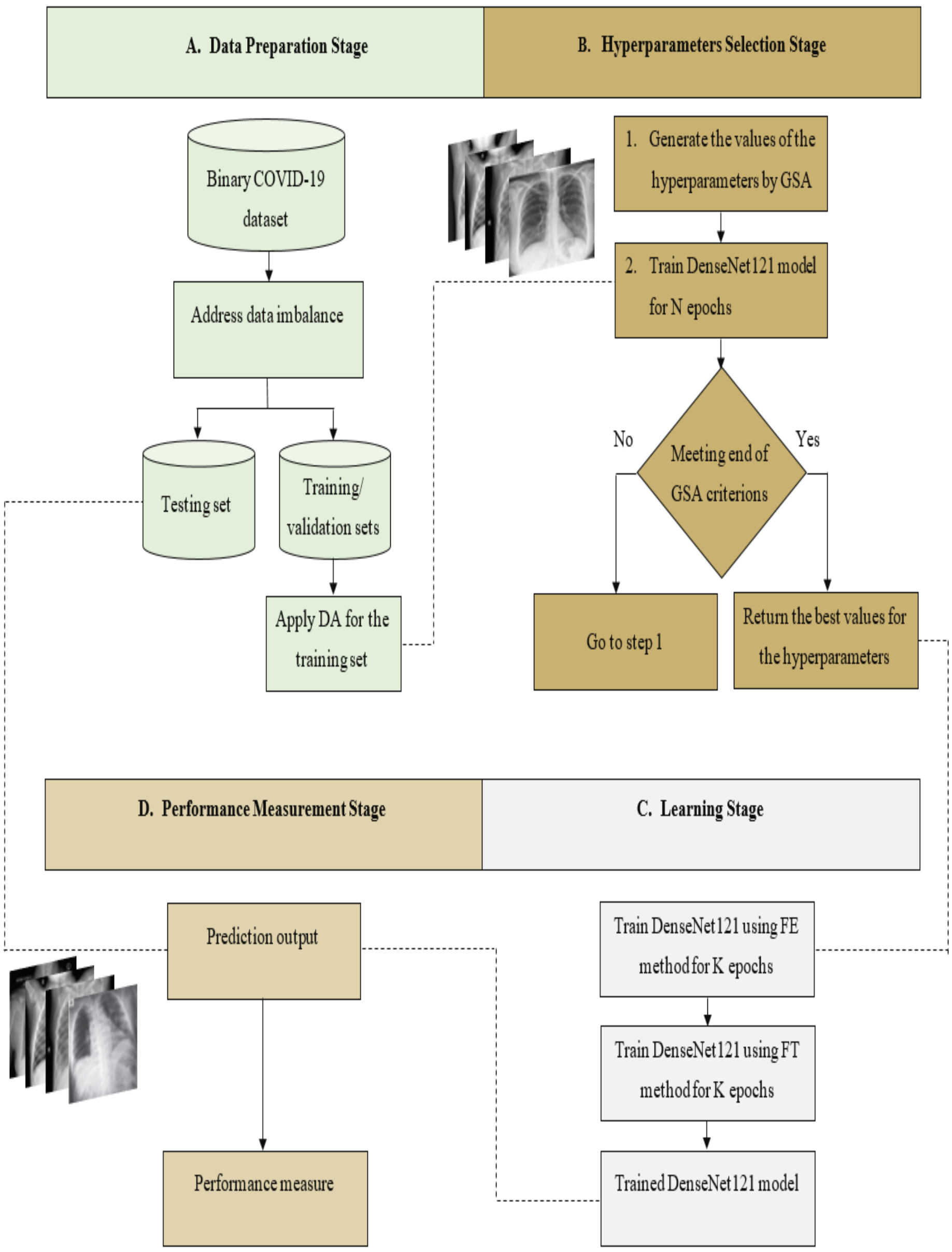

**Figure (4)** the structural form of the proposed approach GSA- DenseNet121 for the binary COVID-19 dataset, DA=Data Augmentation, FE=Feature Extraction, FT=Fine Tuning

***(B) Hyperparameters Selection Stage***

As was previously reviewed, the transfer learning method is taking the same structure of the pre-trained network after making minor changes. The most important minor change is to change the classifier, which requires changing the values of some hyperparameters or adding new hyperparameters. Examples of the hyperparameters that require modification are the batch size, the value of the learning rat, and the number of neurons in the dense layer. The hyperparameters that may add is the rate of the dropout layer. In the proposed approach GSA- DenseNet121, three hyperparameters have been optimized are: the batch size, the rate of the newly added dropout layer, and the number of the neurons of the first dense layer. Therefore, the search space is three-dimensional and each point in the space represents a mixture of these three hyperparameters.

***(C) Learning Stage***

The feature extraction and fine-tuning techniques are utilized to prepare the DenseNet121 architecture to learn from the binary COVID-19 dataset. In the feature extraction, the convolutional base is kept unchanged, whereas, the original classifier base is replaced by a new one that fits the binary COVID-19 dataset. The new classifier consists of four stacked layers and they are a flatten layer, and two dense layers separated by a new dropout layer. The number of neurons in the first dense layer that use RELU as an activation function and the rate of the dropout layer are determined by GSA, and the second dense layer has one neuron with a sigmoid function. After trained the new classifier for some of epochs, the fine-tuning is configured by retraining the last two blocks of the convolutional base of the DenseNet121 with the newly added classifier simultaneously.

### *(D) Performance Measurement Stage*

At this phase, the proposed approach is evaluated. Six measures are utilized to evaluate the proposed approach, namely accuracy, precision, recall, F1 score, and confusion matrix, which are discussed before.

## (5) Experiment Results and Discussion

This section presents and analyzes the results obtained through the proposed approach described in detail in Section 4. All the proposed approach procedures have been implemented using Python with Keras [37]. Keras is a high-level neural network API, written in Python and capable of running on top of TensorFlow, CNTK, or Theano. It was developed for rapid use and the ability to conduct several experiments and get results as quickly as possible and the lowest delay, which helps to conduct good research. To present the results more clearly, they are divided into several sections as follows.

### *5.1 Conducting the Data Augmentation Techniques*

The Keras ImageDataGenerator was utilized to implement augmentation techniques to increase the number of images of the training set of the binary COVID-19 dataset. The data augmentation techniques used and the range used for each technique are listed in Table 1. Figure 5 illustrates some of the images obtained by applying augmentation techniques to one image from each category.

**Table 1.** The augmentation techniques and the range of each technique

| Augmentation Technique | Range |
|---|---|
| Shearing | 0.2 |
| Zooming | 0.3 |
| Width shift | 0.4 |
| Height shift | 0.4 |
| Rotation | 15 |
| Brightness | [0.5, 1.5] |
| Featurewise center | True |
| Featurewise standard deviation normalization | True |
| Fill mode | Reflect |
| Vertical flip | True |
| Horizontal flip | True |

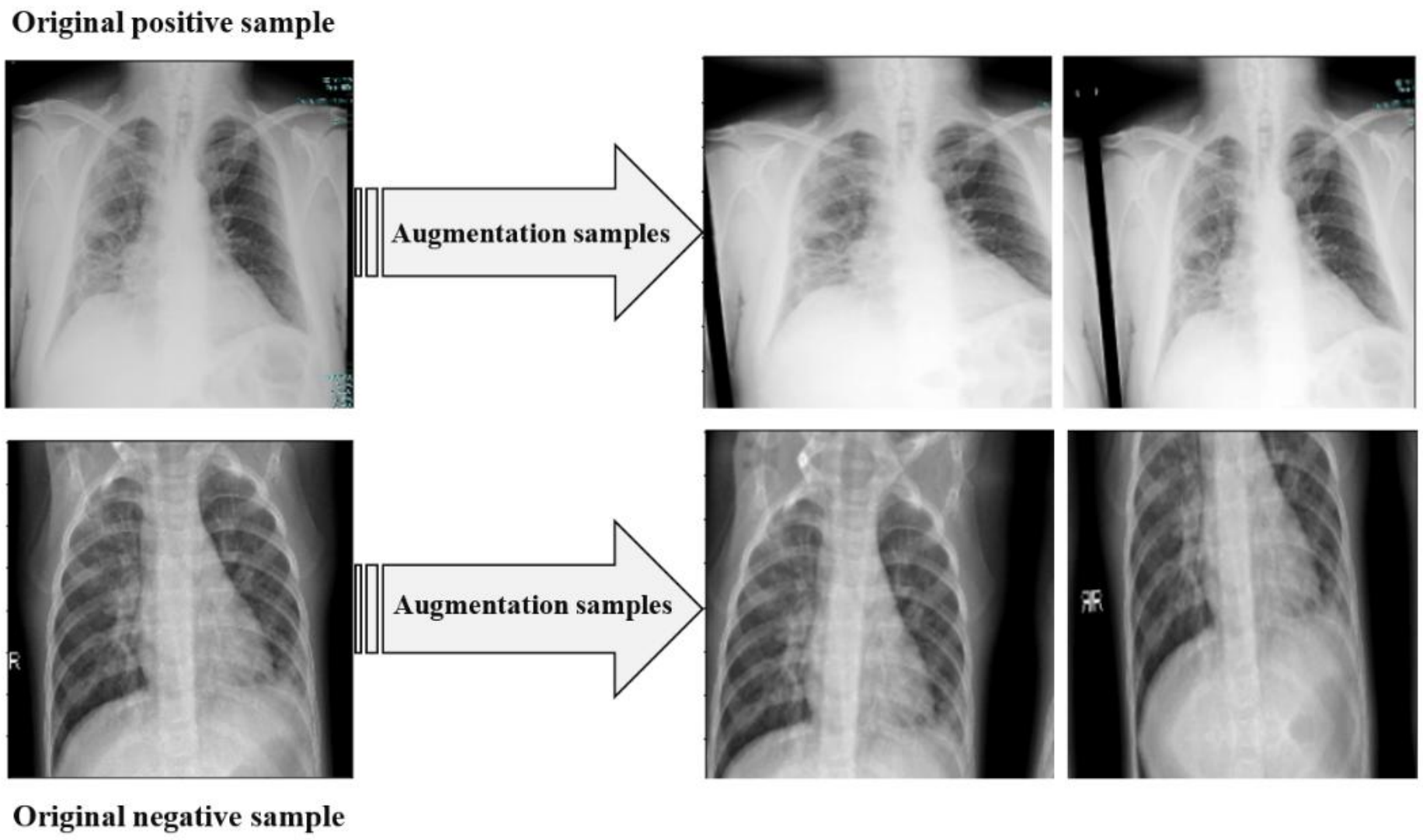

**Figure (5)** Some samples generated from applying various augmentation techniques

### *5.2 Setting up the GSA for the Hyperparameters Selection Stage*

The search space for hyperparameters whose values are to be set by the GSA was bounded as follows: the searching range of batch size was bounded by [1,64], and the searching space of the dropout rate is bounded by [0.1,0.9]. While the searching space of the number of neurons is bounded by [5,500] as listed in Table 2. The values for the GSA parameters were randomly specified, where the maximum number of iteration and population size set to 15 and 20, respectively as listed in Table 2.  While the number of DenseNet121 training epochs was chosen by experimenting with more than one value. Through the experiment, it was concluded that when using a number of epochs over ten, the training process for each of the GSA takes exponential time. While when using less than ten epochs, the results of the DenseNet121 were not sufficiently accurate Therefore, the number of epochs used to train the DenseNet121 was set to ten epochs. The goal of using the GSA is to reduce the loss rate of the validation set as much as possible. The suitability of the proposed GSA solutions is evaluated based on the achieved loss rate in the validation set using these proposed solutions after ten network training periods.

**Table 2.** Preparation of the required parameter values for GSA in conjunction with DensNet121

| **Parameter** | **Value** |
|---|---|
| Maximum number of iterations | 15 |
| Population size | 30 |
| Dimension | 3 |
| Batch size bounds | [1,64] |
| Dropout rate bounds | [0.1,0.9] |
| Number of neurons bounds | [50,500] |
| Maximum number of DenseNet121 training epochs | 10 |

After the GSA training was completed, the optimum values for the batch size, dropout rate, and number of neurons of the first dense layer were determined. Table 3 shows the optimal values for the hyperparameters selected by GSA, where the batch size, dropout and the number of neurons are 8, 0.1, 110 respectively.

**Table 3.** Optimal values for the hyperparameters that were determined by both WOA

| Hyperparameters | Optimal values |
|---|---|
| Batch size | 8 |
| Dropout rate | 0.1 |
| Number of neurons of the first dense layer | 110 |

### 5.3 Learning the DenseNet121 Using the Optimized Hyperparameters

At this stage, the DenseNet121 was trained using the optimal values of the hyperparameters chosen by GSA. DenseNet121 architecture was trained on the training set and evaluated on the validation set for K number of epochs. To determine the value of K, several experiments were conducted, and it was found that the DenseNet121 achieved the best results on the validation set around the 30th epoch within the feature extraction method and about the 40th epoch within the fine-tuning method and that no improvement was observed after that. Thus, the value of K was marked to 30 within the feature extraction and 40 within the fine-tuning. To minimize the overfitting, the process of the training was forced to finish before repetition K if no improvement was perceived for seven iterations; this control was made using early stopping [38]. As the COVID-19 dataset used is a binary-class classification problem, the DenseNet121 is compiled with the binary cross-entropy [39]. The Adam optimizer algorithm [40] was used with a constant learning rate =2e-5 within the feature extraction method. Within the fine-tuning method, a step decay schedule [41] was utilized, where the initial learning rate $LR_0 = 1e - 5$, and the value of the learning rate drops by 0.5 every 10 training epochs. The use of a low learning rate in the fine-tuning method is due to the fact that the amount of changes that will occur in this method should be very small so that the features learned from the feature extraction method are not lost.

### 5.4 Measuring the performance of GSA-DenseNet121

This section presents the results of the performance evaluation of the DenseNet121 architecture using hyperparameters values specified by the GSA. The performance of the proposed approach GSA-DenseNet121 was evaluated using accuracy, loss rate, precision, recall, and F1 score. The proposed approach achieved 98% accuracy in the test set, the macro average and weighted average for the precision, recall, and F score were equal as the values for both were 98% as listed in Table 4.

**Table 4.** The performance of the proposed approach GSA-DenseNet121 and the overall performance is calculated using micro and weighted average

| Categories | Precision | Recall | F1 score |
|---|---|---|---|
| Negative | 97% | 100% | 98% |
| Positive | 100% | 97% | 98% |
| **Macro Average** | **98%** | **98%** | **98%** |
| **Weighted Average** | **98%** | **98%** | **98%** |

To find out the number of samples incorrectly classified by the proposed approach GSA-DenseNet121, as well as the number of samples that it was able to classify correctly, the confusion matrix was used as shown in Figure 6. The dark-colored shaded cells of the confusion matrix represent samples that were correctly categorized in each category, while light-colored shaded cells represent incorrectly categorized samples in each category.

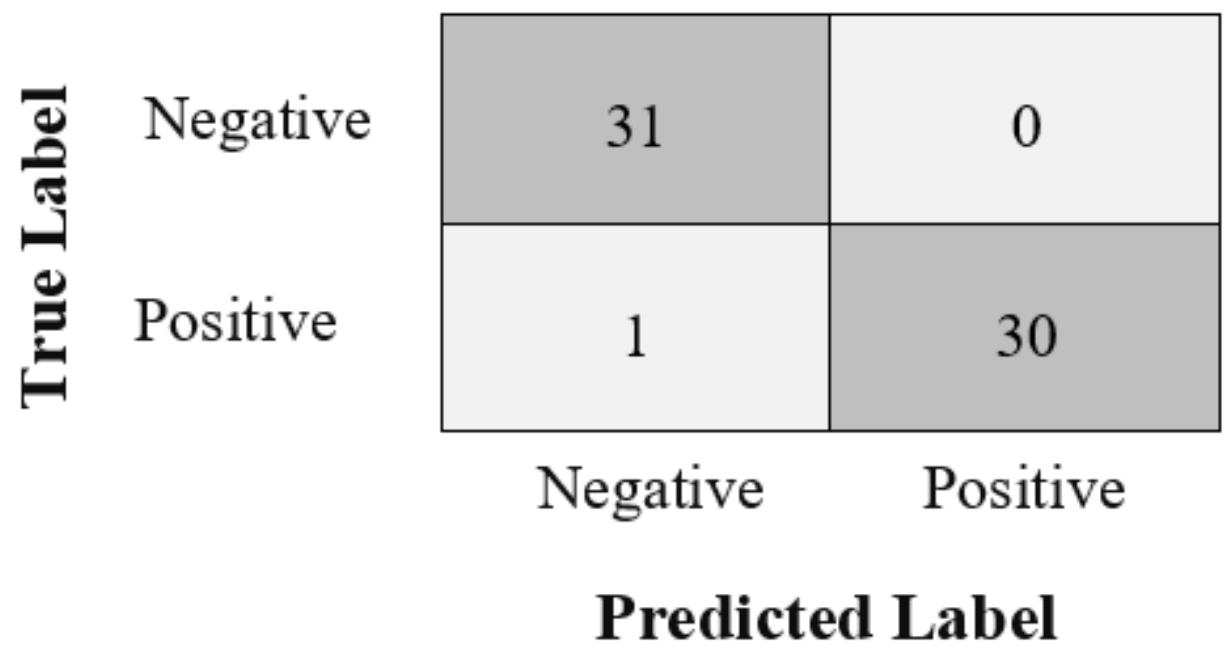

**Figure (6)** the confusion matrix obtained from the proposed approach GSA-DenseNet121-COVID-19

As shown in the confusion matrix in Figure 6, the proposed approach GSA-DenseNet121-COVID-19 was incorrect in classifying only one sample from the test set, while it succeeded in classifying all other samples. Figure 7 shows the sample incorrectly categorized by the proposed approach, as well as some samples for which the proposed approach was successful in categorization.

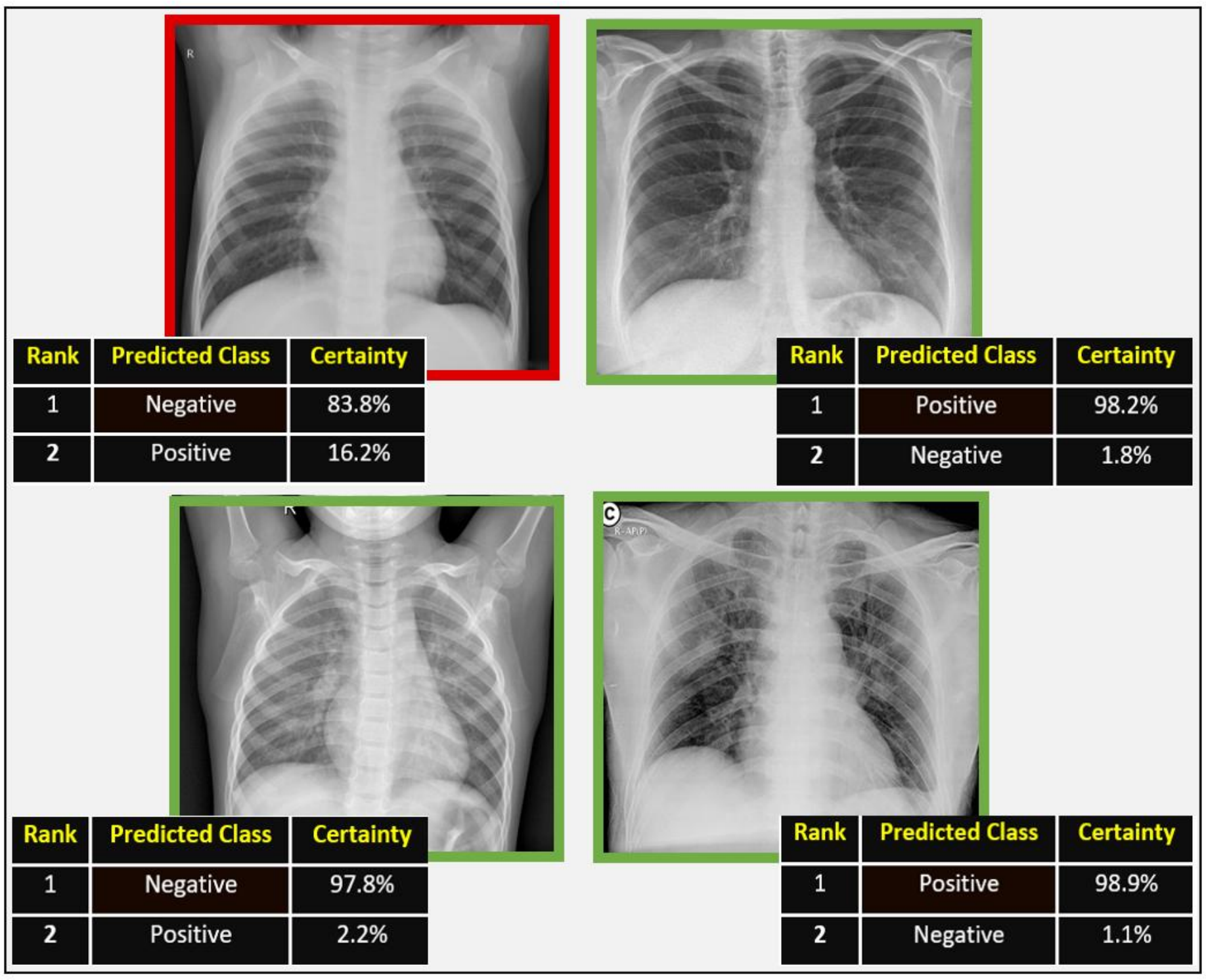

**Figure (7)** some of the correctly labeled samples (enclosed in a green rectangle) and the incorrectly labeled sample (enclosed in a red rectangle) of the test set

To ensure the effectiveness of the GSA in determining optimum values for the hyperparameters of the DenseNet121 architecture that can achieve the highest level of accuracy. It has been compared with the Social Ski (SSD) algorithm [42]. For a fair comparison between GSA and SSD algorithm, the values of the SSD algorithm parameters have been set with the same values that have been set for the GSA parameters as shown in Table 2. After the SSD algorithm has completed the training process, the batch size value was set to 6, while the number of neurons and the dropout rate were set to 220 and 0.71, respectively. The results showed that the GSA is more suitable for pairing with the DenseNet121 to classify the binary COVID-19 dataset, The GSA was able to choose better values for the hyperparameters of DenseNet121 architecture, which in turn made this architecture achieve a higher accuracy ratio. Where the approach SSD-DenseNet121 achieved an accuracy rate of 94% on the test set. As well as the macro average for the precision, recall, F score of the SSD-DenseNet121 were equal as the values for both were 94% as listed in Table 5.

To ensure the performance of the proposed approach GSA-DenseNet121 as a whole, it was compared with the Inception-v3 architecture based on the method of manual search. In this experiment, the values for the VGG16 hyperparameters were randomly chosen, where the values of batch size, dropout rate, and number of neurons were set to 16, 0.5, 250, respectively. This comparison results show the superiority of the proposed approach GSA-DenseNet121-COVID-19 over the Inception-v3 architecture based on the manual search method. The results of the accuracy and the macro average for the precision, recall and F1 score for the inception-v3 architecture are 95%, 96%, 95%, 95%, respectively as shown in Table 5.

**Table 5.** A comparison between the performance of the proposed approach GSA-DenseNet121 and the performance of the SSD-GSA approach. MA-Precision=macro average of precision, MA-Recall= macro average of recall, and MA-F score= macro average of F-score.

| **Comparison items** | **Proposed Approach** | **SSD-DenseNet121** | **Inception-v3** |
|---|---|---|---|
| Batch size | 8 | 6 | 10 |
| Dropout rate | 0.1 | 0.71 | 0.5 |
| Number of neurons | 110 | 220 | 150 |
| Accuracy | 98% | 94% | 95% |
| MA-Precision | 98% | 94% | 96% |
| MA-Recall | 98% | 94% | 95% |
| MA-F Score | 98% | 94% | 95% |

## 6 Conclusions

This paper proposes an approach called GSA-DenseNet121 that can be used to diagnose COVID-19 cases through chest x-ray images. The proposed GSA-DenseNet121-COVID-19 approach consists of four main stages are (1) data preparation stage, (2) the hyperparameters selection stage, (3) the learning stage, (4) the performance measurement stage. In the first stage, the binary COVID-19 dataset was handled from the imbalance and then divided into three sets, namely training set, validation set, and test set. After increasing the number of samples of the training set in the first stage using different data augmentation techniques, it was used in the second stage with the validation set. In the second stage, GSA is used to optimize some of the hyperparameters in the CNN architecture used which is called DenseNet121. In the third stage, DenseNet121 was completely trained using the values of the hyperparameters that were identified in the previous stage which in turn helped this architecture to diagnose 98% of the cases in the test set in the fourth stage. The proposed approach was compared with more than another approach, and the result of the comparison showed the effectiveness of the proposed approach in diagnosing the new virus called COVID-19.